\documentclass[%
 reprint, linenumbers,
superscriptaddress,
 amsmath,amssymb,
 aps, physrev,
]{revtex4-2}

\usepackage{graphicx}
\usepackage{dcolumn}
\usepackage{bm}
\usepackage{xcolor}
\usepackage{algorithm}
\usepackage{algpseudocode}
\usepackage{hyperref}
\begin{document}

\preprint{APS/123-QED}

\title{Plenoptic imaging of particle interactions in scintillation detectors}%

\author{Xiang Dai}
\email{xidai@ucsd.edu}
\affiliation{Department of Electrical and Computer Engineering, UC San Diego, La Jolla, CA, 92093, USA}
\author{Chi-Jui Ho}
\affiliation{Department of Electrical and Computer Engineering, UC San Diego, La Jolla, CA, 92093, USA}
\author{Kevin Tandi}
\affiliation{Department of Electrical and Computer Engineering, UC San Diego, La Jolla, CA, 92093, USA}
\author{Chang Lee}
\affiliation{Lawrence Livermore National Laboratory, Livermore, CA 94550, USA}
\author{Alex Bocchieri}
\affiliation{Department of Computer Sciences, University of Wisconsin-Madison, Madison, WI, 53706, USA}
\author{David Parra}
\affiliation{Department of Computer Sciences, University of Wisconsin-Madison, Madison, WI, 53706, USA}
\author{Forrest Peterson}
\affiliation{Department of Electrical and Computer Engineering, University of Wisconsin-Madison, Madison, WI, 53706, USA}

\author{Talha Sultan}
\affiliation{Department of Electrical and Computer Engineering, University of Wisconsin-Madison, Madison, WI, 53706, USA}
\author{Felicia Sutanto}
\email{sutanto2@llnl.gov}
\thanks{Corresponding author.}
\affiliation{Lawrence Livermore National Laboratory, Livermore, CA 94550, USA}

\author{Andreas Velten}
\affiliation{Department of Biostatistics and Medical Informatics, University of Wisconsin-Madison,
Madison, WI 53706, USA}
\affiliation{Department of Electrical and Computer Engineering, University of Wisconsin-Madison, Madison, WI, 53706, USA}

\author{Jingke Xu}
\affiliation{Lawrence Livermore National Laboratory, Livermore, CA 94550, USA}

\author{Nicholas Antipa}
\affiliation{Department of Electrical and Computer Engineering, UC San Diego, La Jolla, CA, 92093, USA}

\date{\today}

\begin{abstract}
Accurate 3D localization of radiation interactions in scintillation detectors is critical for applications in nuclear and particle physics, safeguards, and medical imaging, yet remains challenging in light-starved regimes with limited photon statistics. 
We designed a multifocal plenoptic imaging system that balances photon collection and spatial encoding to overcome weaknesses of conventional unifocal plenoptic systems under photon-limited conditions. 
A Cramér--Rao lower bound analysis guides the multi-focal microlens array design to achieve high effective numerical apertures for millimeter-scale position resolutions within a single-volume scintillator. 
We constructed a prototype and calibrated its response with a tunable light source. The recorded images were denoised to create realistic data with $\mathcal{O}$(100) detected photons, the reconstruction of which 
demonstrated $\mathcal{O}$(1)~mm 3D localization for single-vertex events and slightly larger errors for double-vertex events. This work establishes a promising approach for photon-efficient 3D reconstruction in light-starved environments and provides a foundation for future exploration of enhanced multi-scattering event reconstruction in single-volume detectors.

\end{abstract}

\maketitle

\section{Introduction}

Scintillation detectors convert energy deposits by ionizing radiation into photon signals that can be efficiently collected using mature and cost-effective sensors and electronics, allowing for the interaction time, energy deposition, and event topology to be reliably reconstructed. 
In addition, some scintillators produce scintillation time profiles that depend on the interaction type, enabling particle identification through pulse shape discrimination (PSD). 
Non-crystalline scintillators are scalable and can be fabricated in a wide range of sizes and shapes to accommodate the needs of different applications. 
Therefore, scintillation detectors are widely used in particle physics experiments, nuclear engineering, nuclear medicine, and clinical imaging. 

Accurate 3D localization of energy depositions in the scintillation volume can greatly enhance the detector performance.
For instance, precise event positioning enables fiducialization to suppress external backgrounds~\cite{chooz_2003,kamland_2005,songs_2007, borexino_2015, solid_2021, prospect_2019, prospect_2021, dazeley_2025,lux_2019_PRL,lux_2019_PRD}, improve energy and timing reconstruction~\cite{smy_2008}, and provide access to event topology to support particle identification~\cite{liquido_2023,liquido_2025}. 
These capabilities are critical for state-of-the-art neutrino and dark matter experiments as well as deployable radiation detectors. 

Position reconstruction in single-volume scintillator detectors is usually achieved through photon time of flight (TOF) and/or the spatial gradient of photon intensity~\cite{smy_2008}, using the fact that isotropically emitted photons are more likely to reach nearby sensors at an earlier time and at a higher flux than for faraway sensors. 
Single-volume detector modules can be further combined into larger detectors, where each subvolume is optically segmented from others and the identity of the segment also encodes approximate event position information. 
These techniques and their combinations have enabled position resolutions ranging from the millimeter scale to the tens-of-centimeters scale, depending on event energy, detector size, design, and scintillator material~\cite{sandd_2021,prospect_2021,smy_2008,CeBr3_mmScale}. 

Nuclear safeguards and nonproliferation missions can benefit from radiation imaging technologies that combine a compact geometric footprint with scalable active material volume, supporting a wide range of applications, including standoff detection, nuclear material characterization, detection of emerging warhead designs, waste and hold-up measurements, contamination remediation, and warhead counting. 
Today, compact gamma imagers are commercially available but mainly rely on crystalline detector volumes~\cite{vetter_2007,vetter_2018,h3d_2018,h3d_2019,polaris_2021}, which can hinder scalability. 
Compact neutron imagers, on the other hand, remain an active area of research, further limited by degraded event-reconstruction performance with low detectable energies from neutron interactions~\cite{braverman2018single,sandd_2019,sandd_2021} and the lack of adequate high-density readout systems~\cite{wu_2023,li_2023}. 
The scalability challenge of neutron and gamma imagers can be addressed with the use of non-crystalline scintillators, which, however, typically have lower light yield than crystalline scintillators and have not allowed the desired millimeter-scale 3D position resolution~\cite{braverman2018single} to be achieved with $\mathcal{O}$(100) detected photons.

Two primary design approaches have been explored for the development of compact and scalable position-sensitive radiation imagers: optically segmented detectors~\cite{Weinfurther_2018,manfredi2020single,galindo2021design,wu_2023,li_2023} and monolithic (single-volume) detectors~\cite{folsom2020compact,ziock2019real,braverman2018single}.
In this optically segmented design, the scintillator is divided into discrete 2D segments~\cite{galindo2021design,manfredi2020single}, where lateral interaction coordinates are inferred from the segment identity while the axial position is estimated via timing or amplitude differences in signals collected by coupled photosensors at the ends. Consequently, lateral resolution is governed by the segment cross-section and can be refined through finer segmentation if suitably sized sensors and high-density electronics are available. Conversely, the axial resolution is fundamentally limited by scintillation speed and photon statistics, with only centimeter-scale precision demonstrated to date in plastic scintillators for 1~MeV energy depositions~\cite{sandd_2019,sandd_2021}. Furthermore, the photosensors coupled to the ends of the scintillators need to provide both high temporal resolution and robust charge linearity, which, alongside the high channel density requirement for spatial precision, pose a serious challenge to the readout system.

The single-volume approach, in principle, could simplify detector design and provide a more uniform angular response. It also allows more scintillator faces to be accessed so spatial information and timing information could be separately collected, thereby relaxing the requirements on the sensor readout system. However, these potential benefits can only be realized if challenges of position reconstruction in single-volume detectors are adequately addressed. 
To date, techniques such as centroiding~\cite{braverman2018single,harvey2021development} and coded-aperture (amplitude mask) methods~\cite{folsom2020compact,ziock2019real} have not demonstrated mm-scale 3D spatial resolution in organic scintillators, for either single-vertex or multi-vertex particle interactions. 

We aim to achieve millimeter-scale 3D localization in a large single-volume scintillation detector under light-starved conditions by investigating the plenoptic imaging approach~\cite{renNg_2005}. 
The plenoptic system can encode both spatial and angular information of detected photons through a multi-focal microlens array (MLA), and provide depth-sensitive measurements that are well-suited for 3D reconstruction. 
Prior attempts of plenoptic imaging in scintillation detectors have demonstrated the potential of this technology~\cite{dalmasson,dieminger}, but their adoption of conventional plenoptic lens designs resulted in low light collection efficiencies or weak depth encoding. 
In this work, we systematically evaluate the trade-off between light collection and spatial encoding and use the Cramér--Rao lower bound (CRLB), derived from Fisher information, to achieve an improved optical design. 

\begin{figure}
    \centering
    \includegraphics[width=1.0\linewidth]{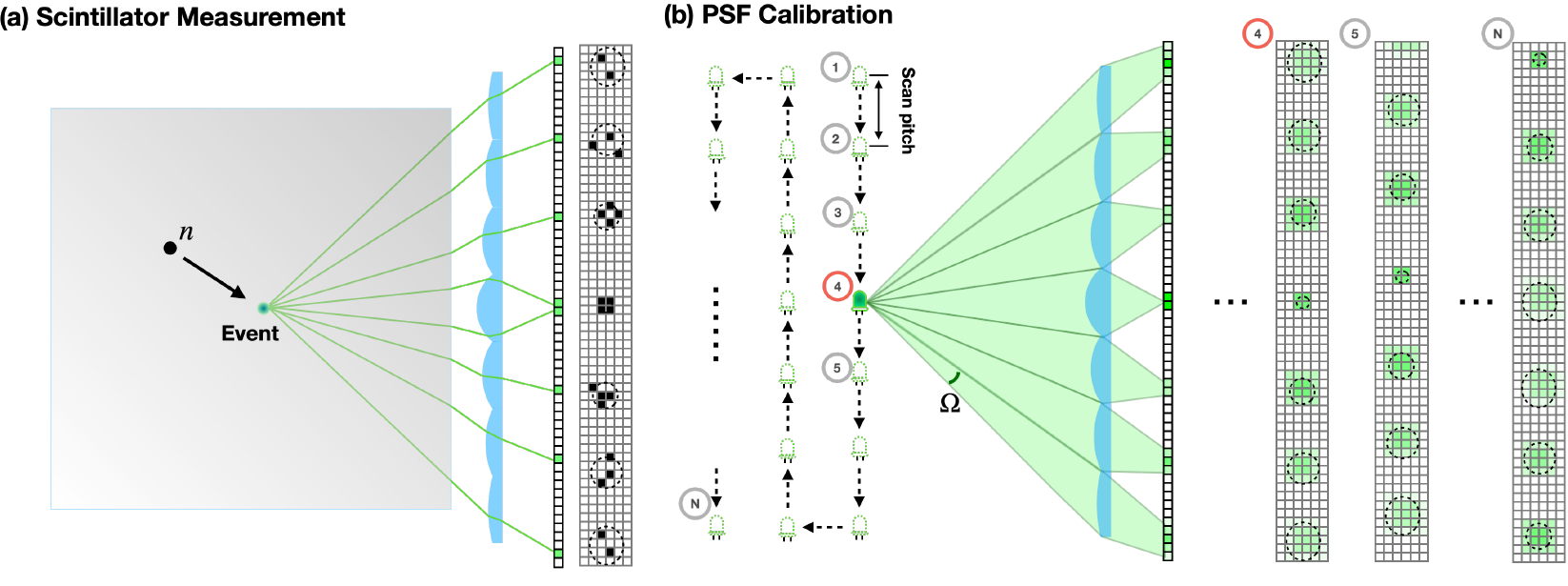}
    \caption{PSF calibration and measurement with a multifocal lens array. \textbf{(a)} Scintillator measurement: A scintillation event emits photons isotropically, which are collected by the multifocal lens array and projected onto the single-photon-sensitive sensor, resulting in a binary photon detection pattern.\textbf{(b)} A point-like LED source is scanned throughout the object volume along the $(x, y, z)$ coordinates. The dashed outlines denote the corresponding circles of confusion for different depths. By sampling the LED across axial positions, a 3D PSF dictionary is constructed. $\Omega$ denotes the solid angle subtended by each lenslet, which governs photon collection efficiency.}
    \label{Fig.1:system_overview}
\end{figure}

Guided by this analysis, we design the Plenoptic Radiation Imager for Spectroscopic and Directional Mapping (PRISM) system, a compact plenoptic architecture that can be positioned close to a single-volume scintillator to prioritize light collection while using a multi-focal MLA to encode depth information. 
The design assumes a low-noise, single-photon-sensitive sensor, such as a single-photon avalanche diode (SPAD) array or a high-rate picosecond photodetector (HRPPD), to suit operation in photon-limited regimes. 
The sensor is assumed to operate in a short-gate binary measurement mode, which suppresses read noise and dark-count to negligible levels, yielding measurements dominated by photon statistics.
We further construct a prototype and experimentally characterize its point-spread functions (PSFs) under high-light-intensity conditions. 
The acquired PSFs are then denoised to produce test images in the light-starved regime using the single-photon-sensitive sensor model to test reconstruction accuracies. 
The process is illustrated in Fig.~\ref{Fig.1:system_overview}, where different source locations produce distinct sensor patterns, allowing for the 3D event position to be reconstructed. 
We present the design, construction, and calibration of the PRISM prototype, together with single-vertex reconstruction in the photon-limited regime using sparse recovery algorithms. 
We also provide a preliminary assessment of its potential for double-vertex reconstruction.

This paper is organized as follows.
Section~\ref{sec:theory} introduces candidate plenoptic system designs and develops the Fisher information framework to evaluate theoretical limits on spatial reconstruction performance for the studied optical systems. 
Section~\ref{sec:experiment} details the prototype construction, system calibration, data acquisition, and the approach used to convert high-light-intensity measurements into data-driven predictions under light-starved conditions.
Section~\ref{sec:model_and_alg} 
describes the forward model with a binary single-photon-sensitive sensor measurement mode and the reconstruction method. 
Section~\ref{sec:results} evaluates the system's performance in reconstructing both single-vertex and double-vertex interactions. 
Section~\ref{sec:discussion} discusses these results in the context of existing single-volume reconstruction approaches, including centroiding and coded-aperture methods.
Finally, Sec.~\ref{sec:conclusion} summarizes the findings and outlines directions for future work.

\section{Theoretical Performance Predictions} 
\label{sec:theory}

In this section, we evaluate the theoretical performance of a few optical designs in encoding depth information for sparse scintillation events in the light-starved regime. 
The scintillation volume is assumed to be $5 \times 5 \times 5~\mathrm{cm}^3$, and all optical systems share a working distance of 5~mm from one face of the scintillator and an imaging distance of 18.5~mm. The small distances are chosen to prioritize the system's light collection efficiency, which is critical for light-starved imaging, and to overcome the extremely low photon statistics in conventional imaging approaches as demonstrated in reference \cite{dieminger}. 
This shared geometry isolates the effect of optical architecture from differences in object volume or sensor placement.

\subsection{Optical design options}
\label{sec:optics}

\begin{figure}
    \centering
    \includegraphics[width=1\linewidth]{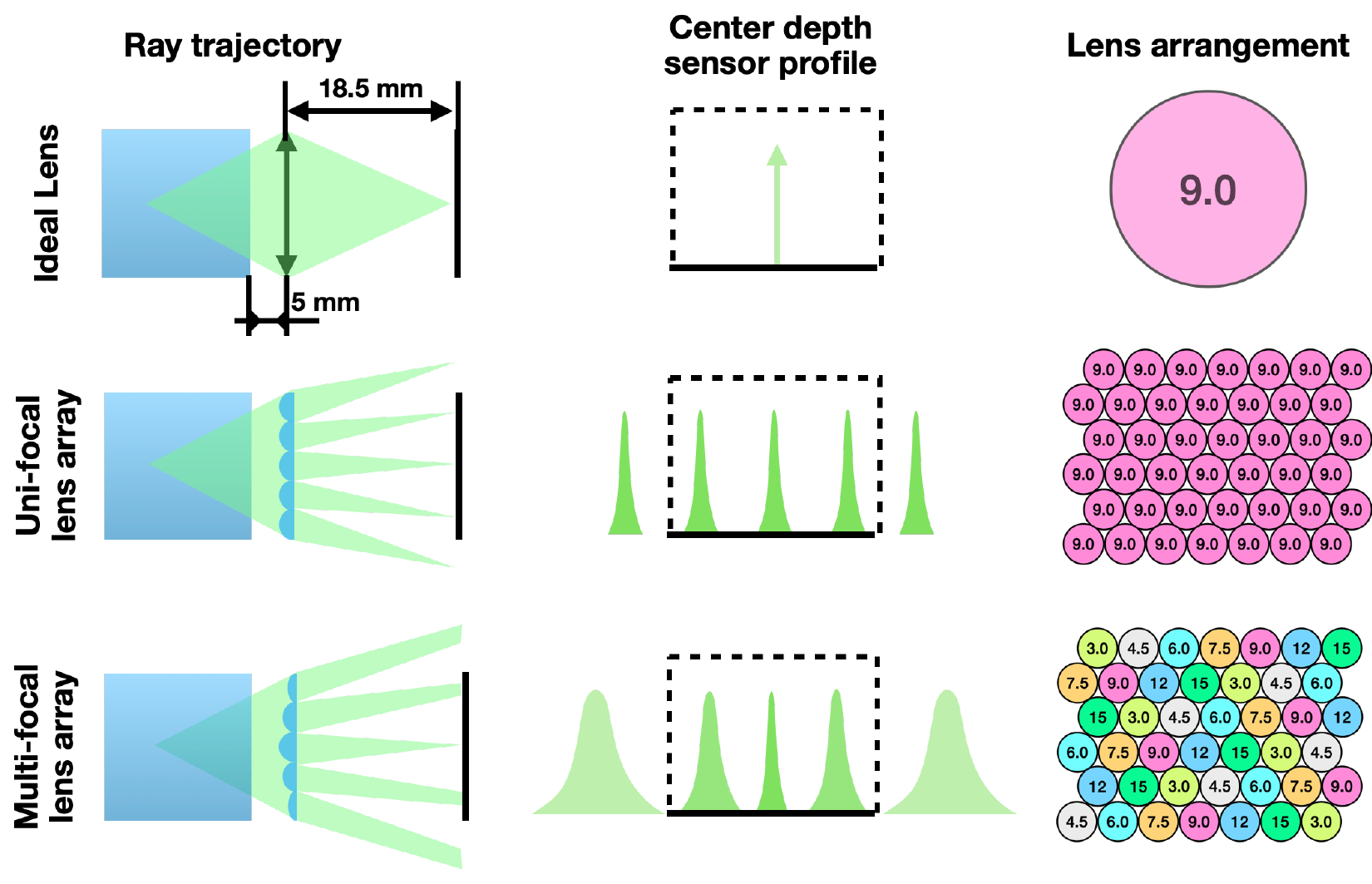}
    \caption{Imaging configurations and corresponding responses. \textbf{Top:} Ideal single-lens system. \textbf{Middle:} Uni-focal lens array. \textbf{Bottom:} Multi-focal lens array. Left column shows ray trajectories, and the center column shows the corresponding sensor responses along depth, with dashed boxes indicating the sensor limitation. The right column shows the lens layouts; numbers denote the focal lengths (mm) of individual lenslets.
}
    \label{fig:optical_designs}
\end{figure}

The three configurations considered in this work are illustrated in Fig.~\ref{fig:optical_designs}. 
The first row illustrates the ideal single-lens configuration. 
To achieve a high photon collection efficiency, a small $f$-number is required, corresponding to a large aperture relative to the focal length. 
This requirement is difficult to realize with a single physical element, and a practical implementation would require multiple large optical elements, increasing the working distance and making integration into compact detector geometries difficult.
For simplicity, this design is modeled as an ideal lens and serves primarily as a theoretical reference for performance evaluation. 
A sharp image may be formed with a single lens when the thin lens equation is satisfied, and depth information of the light source is encoded in the resulting defocus blur. 

The same photon-collection constraint also motivates removing the main lens used in conventional plenoptic architectures.
To improve compactness while maintaining high photon collection efficiency, we next consider a uni-focal MLA, as shown in the second row of Fig.~\ref{fig:optical_designs}. 
In this architecture, the total collection aperture is distributed across multiple lenslets with identical diameters and focal lengths, allowing the array to be positioned close to the scintillation volume. 
This arrangement maintains a large photon collection solid angle while substantially reducing the overall system form factor. 
Unlike a conventional plenoptic camera, this configuration does not use a main lens, so emissions from different spatial locations can produce multiplexed measurements on the sensor.
However, in the single- and double-scattering regimes considered in this work, allowing the resulting multiplexing to be resolved by the sparse reconstruction framework. 
The primary limitation of the uni-focal MLA is that all lenslets share the same focal length and similar defocus behaviors, which restricts the available depth diversity and therefore limits the axial sensitivity range.

To increase depth diversity, we next investigate a multi-focal MLA architecture in which lenslets with identical diameters but different focal lengths are incorporated within the same array, as illustrated in the third row of Fig.~\ref{fig:optical_designs}. 
The common lenslet diameter simplifies assembly, while the variation in focal length introduces focal diversity in the PSF across the volume. 
This diversity, in principle, should make measurements from different depths more distinguishable, thereby improving depth encoding and 3D localization performance.

In the subsequent sections, we evaluate the localization performance against a uni-focal microlens-array baseline, while retaining the single-lens configuration as an ideal thin-lens reference. For both configurations, we consider three focal lengths, $f \in \{3, 9, 15\}$ mm, chosen such that the nominal focal planes lie near the front, middle, and rear of the object volume for an image distance of 18.5 mm. The single-lens reference is simulated using the thin-lens model.

\begin{figure*}
    \centering
    \includegraphics[width=\linewidth]{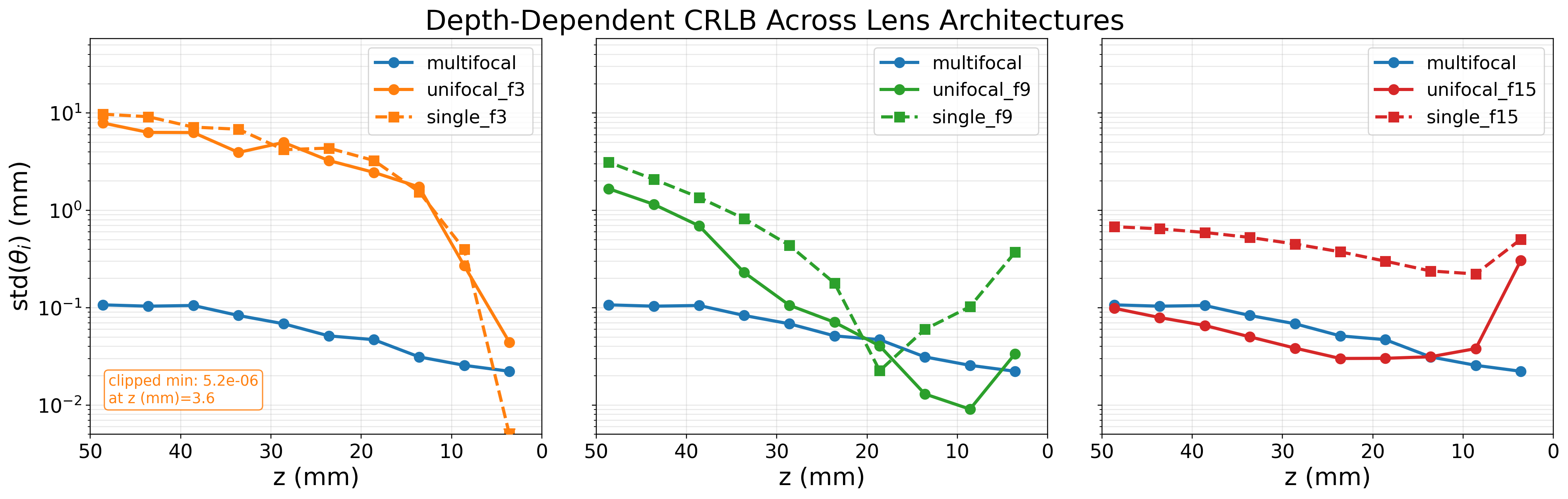}
    \caption{Comparison of Fisher Information across different systems and focal lengths. The plots compare the proposed multifocal system, a baseline unifocal system, and a reference single-lens system. Figures from left to right correspond to focal lengths of 3, 9, and 15 mm. The depth of the isotropic light source ranges from 0 to 50 mm from the MLA. }
    \label{fig:crlb}
\end{figure*}

\subsection{Localization Sensitivity}\label{sec:fisher_intro}
To evaluate the localization sensitivity of each optical design, we use the Fisher information matrix (FIM), which quantifies how the expected intensity measurement changes in response to small shifts in the source position $\theta = [x,y,z]^T$.
The utility of the FIM is established by the Cram'{e}r-Rao inequality \cite{cramer1999mathematical, steven1993fundamentals}, which lower-bounds (CRLB) the uncertainty achievable by any unbiased localization estimator. Specifically, the variance of each estimated spatial coordinate is bounded by the corresponding diagonal element of the inverse FIM $I(\theta)$. For the lateral coordinate $x$, this bound is expressed as:

\begin{equation}
\label{eq:var}
\mathrm{Var}(\hat{x}) \ge \left[I^{-1}(\theta)\right]_{xx} \ge \frac{1}{I_{xx}(\theta)}
\end{equation}
with analogous bounds holding for the $y$ and $z$ coordinates.

These lower bounds benchmark the system precision limit. Given a shot-noise-limited intensity measurements $\mathbf{b}(\theta)$, the Fisher information is given by \cite{chao2016fisher}:

\begin{equation}
\label{eq:fisher_split}
\begin{split}
I(\theta) = \sum_{i} & \left( \frac{\partial b_{i}(\theta)}{\partial \theta} \right)^{T} \left( \frac{\partial b_{i}(\theta)}{\partial \theta} \right) \cdot \frac{1}{b_{i}(\theta)}, \theta \in \Theta ,
\end{split}
\end{equation}

\noindent where $b_{i}(\theta)$ is the measurement intensity at pixel $(i)$. 
To calculate these spatial gradients in practice, we introduce a differentiable ray-tracing framework in the next section.

To evaluate the Fisher Information in Eq.~\ref{eq:fisher_split}, we use a differentiable ray-tracing framework \cite{wang2022differentiable, ho2025differentiable} implemented in PyTorch \cite{paszke2019pytorch}. 
The framework models ray emission, MLA refraction, and geometric transport to the sensor plane, producing densely sampled ray distributions and the corresponding measurement $\mathbf{b}(\theta)$.

To maintain the differentiability of $\mathbf{b}(\theta)$, we model the discrete sensor response through bilinear interpolation, which smoothly maps continuous ray intersections onto the discrete pixel grid. 
Consequently, this computational pipeline ensures that the model captures the sensitivity of sensor measurements to infinitesimal perturbations in $\mathbf{\theta}$, enabling the direct computation of the partial derivatives $\frac{\partial \mathbf{b}(\theta)}{\partial \theta}$. Through Eq.~\ref{eq:var} and~\ref{eq:fisher_split}, we compute the directional information terms and the corresponding CRLB standard deviation bounds across the scintillator volume, and compare the depth-encoding performance under different lens configurations.

\subsection{CRLB on different MLA designs}
\label{Sec:CLRB_results}

We evaluate the CRLB across a range of depths by benchmarking our proposed multifocal design against the two configurations introduced in Sec.~\ref{sec:optics}: the unifocal baseline and the single-lens reference. The simulation utilizes an isotropic light source and the ray-tracing engine described in Sec. \ref{sec:fisher_intro}. These results, visualized in Fig. \ref{fig:crlb}, illustrate the joint effects of energy concentration and energy amount on localization precision.

Regarding energy concentration, an in-focus spot naturally yields lower localization uncertainty than an out-of-focus spot, yielding the inherent limitations of the unifocal baselines and single-lens reference across the tested volume. For $f=3$~mm, the narrow depth-of-field at small $z$ causes both the unifocal and single-lens systems to exhibit severe localization uncertainty. Similarly, the $f=9$~mm system maintains low uncertainty only within a narrow band around its focal depth. While the $f=15$~mm unifocal design achieves a low CRLB at deeper planes, its precision degrades rapidly at closer distances. Notably, at $z = 3.5$~mm, our multifocal design achieves a $35\times$ lower CRLB than the $f=15$~mm baseline. These results confirm that while unifocal systems are restricted to narrow depth-of-field peaks, the proposed multifocal architecture maintains a consistently low CRLB across the entire volume by encoding depth information into highly distinctive patterns.

Regarding the amount of energy, we observe that the CRLB generally decreases as depth ($z$) decreases. This trend is driven by the depth-dependent geometric efficiency of an isotropic source emitting 6280 rays; a closer distance to the MLA is associated with a larger solid angle, allowing the sensor to capture more photons and collect richer information. Specifically, the expected photon counts vary with depth, ranging from 60 to 500 photons. Notably, the $f=15$~mm single-lens design deviates from this trend. Despite the significantly increased photon statistics at close proximity, its performance rapidly deteriorates because the severe spatial divergence of the out-of-focus wavefront overpowers the benefits of higher photon counts. In contrast, our proposed multifocal design yields consistently low uncertainty over the entire tested depth range, demonstrating its clear superiority over both the baselines and the reference single-lens systems.

\section{Experimental System and Implementation} \label{sec:experiment}

Based on the theoretical studies discussed in Sec.~\ref{sec:theory}, we designed and constructed a compact multifocal MLA system and calibrated its responses to a point light source as a function of source positions. In this section, we discuss the detailed design of the PRISM prototype and its calibration. 

\subsection{PRISM system}
\begin{figure}
    \centering
    \includegraphics[width=1.0\linewidth]{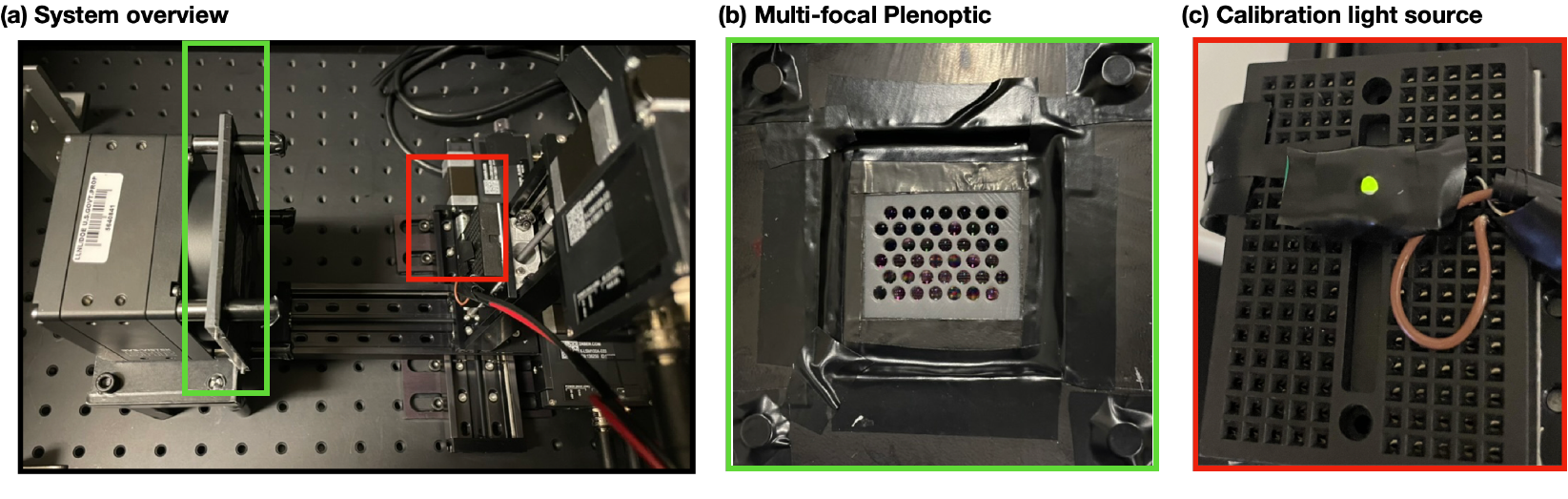}
    \caption{\textbf{(a)} PRISM system overview \textbf{(b)} Green box: Multi-focal plenoptic lens array. \textbf{(c)} Red box: Calibration LED source.
}
    \label{fig:calibration_system}
\end{figure}
The PRISM prototype is shown in Fig.~\ref{fig:calibration_system}(a). 
The multifocal plenoptic lens array was constructed using commercially available 3~mm-diameter plano-convex microlenses (Edmund Optics) with focal lengths of 3, 4.5, 6, 7.5, 9.0, 12, and 15~mm, providing focal-length diversity for depth encoding. 
All lenses were coated with a visible broadband anti-reflection coating with extended UV performance to improve optical transmission.

Due to component availability at the time of assembly, the prototype used 42 lenses arranged in a hexagonal packing pattern. 
The lenses were mounted in a custom holder, shown in Fig.~\ref{fig:calibration_system}(b), and the detailed focal-length arrangement is shown in the bottom row of Fig.~\ref{fig:optical_designs}.
The current prototype provides lateral coverage of approximately $2 \times 3~\mathrm{cm}^2$, primarily limited by the availability of suitable commercial lenses rather than design choices. 
In the following experiments, we characterize the prototype over this accessible field of view within the intended $5 \times 5 \times 5~\mathrm{cm}^3$ localization volume.

The lens holder was fabricated using a Formlabs Form~3+ stereolithography printer with Tough~2000 resin, providing sufficient stiffness for stable optical alignment. 
Each lens was secured using a friction-fit geometry, with an additional retaining lid to constrain the axial position and maintain consistent lens-to-sensor spacing across the array. 
The holder was coupled directly to the image sensor through a cage-system interface, enabling reproducible positioning and reducing the need for manual fine adjustment.

The image sensor was an SVS-SHR661MCX12 camera with an active area of 46.2~mm~$\times$~32.87~mm and a native pixel pitch of 3.45~$\mu$m. 
Throughout the experiments, $2\times2$ pixel binning was applied, resulting in an effective pixel size of 6.9~$\mu$m, to increase the collected signal per effective pixel and to reduce the impact of readout noise.

In this study, system calibration was performed in air using a controllable point source. 
This choice was motivated by practical constraints, as it is impractical to generate calibration scintillation events at known 3D locations inside a scintillator volume.
The resulting calibration captures the measured mapping between 3D source position and sensor response and is used as the data-driven forward model for position reconstruction. 
This calibration framework can also be adopted to reconstruct real scintillation events by incorporating refractive-index effects and boundary interactions introduced by the scintillator material. 
These effects may include refraction at material interfaces, total internal reflection, and photon trapping, which will be studied in future experimental work.

\subsubsection{Calibration light source and control}

The calibration system uses a motorized three-axis translation stage (X-LSM100A-E03, Zaber Technologies) with a travel range of 101.6~mm along each axis, sufficient to cover the full scanning volume. 
The stage provides a positioning accuracy of approximately 35~$\mu$m, enabling reproducible placement of the calibration source.

An LED light source (LG~R975, Osram) with a nominal peak emission wavelength of 570~nm was mounted on the translation stage to serve as a controllable point-like emitter, as shown in Fig.~\ref{fig:calibration_system}(b). 
This wavelength was chosen to approximate the emission spectrum of the intended scintillator, making the calibrated optical response representative of scintillation measurements. 
The LED emission was diffused, and non-emitting surfaces of the LED assembly were masked to suppress stray reflections. 
The calibration setup was enclosed to minimize ambient illumination, and data acquisition was fully automated.


\subsubsection{HDR PSF calibration}\label{sec:hdr}

The imaging system exhibits approximate local shift invariance, meaning that small lateral changes in source position produce PSFs that can be approximated as spatial translations of nearby calibrated PSFs. 
This property reduces the number of calibration measurements, allowing the PSF dictionary to be sampled on a sparse spatial grid.
In our system, calibration is performed over the usable depth range using 10 axial planes with a spacing of 5~mm.
At each depth plane, PSFs are measured on an $11\times11$ lateral grid with a spacing of 5~mm, covering the usable lateral field of view.

In a multifocal plenoptic system, the captured PSFs exhibit a wide dynamic range because both focused and highly defocused responses appear simultaneously on the sensor. 
To preserve both bright and weak PSF features while reducing noise, we apply a high dynamic range (HDR) preprocessing strategy adapted from~\cite{dai2025single}.

For an exposure time $T$ and calibration position $\theta$, the measurement at pixel $(u,v)$ is modeled as

\begin{equation}
b[u,v|\theta,T] = Q\{Tp[u,v|\theta] + \eta_{\theta,T}\},
\end{equation}

where $p[u,v|\theta]$ is the exposure-independent PSF intensity, $Q\{\cdot\}$ denotes sensor clipping and quantization, and $\eta_{\theta,T}$ represents measurement noise.

For unsaturated pixels, the PSF intensity can be estimated from the exposure-normalized measurement as
\begin{equation}
f_{T_j} = \frac{b[u,v|\theta,T_j]}{T_j}.
\end{equation}
At each calibration position, we acquire $m=7$ measurements with different exposure times and collect the exposure-normalized values at each pixel as
\begin{equation}
\mathbf{f} = [f_{T_1}, f_{T_2}, \ldots, f_{T_m}]^\intercal .
\end{equation}
A validity vector $\boldsymbol{\epsilon}$ is calculated to indicate which exposure-normalized measurements are valid, excluding saturated pixels and low-quality measurements.
The HDR PSF estimate is then computed by averaging the valid exposure-normalized measurements:
\begin{equation}
\hat{p}[u,v|\theta] =
\frac{\boldsymbol{\epsilon}^\intercal \mathbf{f}}
{\mathbf{1}^\intercal \boldsymbol{\epsilon}} .
\end{equation}
Pixels are retained only if they have at least three valid exposures, a correlation coefficient greater than 0.95 between measured intensity and exposure time, and acceptable dark-current levels as determined from dark-frame measurements.

Repeating this procedure over all calibration positions produces the HDR PSF dictionary $\mathbf{P}$ used in the reconstruction forward model. 
Each calibrated PSF $\hat{p}[u,v|\theta]$ represents the expected spatial distribution of detected photons produced by an emission at position $\theta$, up to a global photon-count scale.
Simulated measurements for algorithm validation can then be generated by sampling from these calibrated photon-distribution maps using a chosen sensor model.

\section{Reconstruction Algorithm} \label{sec:model_and_alg}

To test the performance of the constructed PRISM prototype for localizing scintillation events, we generate low-light exposure images by sampling a specified number of expected photons from an HDR-calibrated PSF data with a random location, reconstruct their source positions and compare the estimates with the ground truth. These test images, despite being simulations, contain realistic experimental effects such as from system misalignment and optical aberrations, and at the same time can support any chosen sensor response model for sensitivity predictions. This section details the forward model, the reconstruction algorithm and their implementations.

\subsection{Single-photon-sensitive sensor model}\label{sec:SPAD model} 

Although the PRISM prototype uses a CMOS sensor that cannot detect a single photon in a pixel, we assume the acquisition is single-photon-sensitive to evaluate the system's predicted performance for light-starved applications. 
We also assume a finite pixel size with narrow temporal gate ($\sim$ns) to suppress dark counts and readout noise.
Under these assumptions, sensor noise will be negligible, and the response of photon arrivals at pixel $(i)$ can be modeled as a Poisson process.

Let $P_{i}(\theta)$ denote the normalized PSF for a light source located at $\theta=(x,y,z)$, such that: 

\begin{equation}
\label{eq:norm_psf}
\sum_{i} P_{i}(\theta)=1.
\end{equation}

The expected photon rate $\lambda_{i}(\theta)$ at each sensor pixel is then scaled by the expected detected photon count $\alpha$:

\begin{equation}
\label{eq:photon_rate}
\lambda_{i}(\theta) = \alpha\, P_{i}(\theta).
\end{equation}

For a binary photon-counting sensor such as a SPAD array, which is chosen for this study to be conservative, a pixel records whether at least one photon is detected during the temporal gate. 
With time-gating, false-positive detections are assumed to be negligible. Consequently, the binary measurement $y_{i} \in \{0, 1\}$ at each pixel is modeled as a Bernoulli random variable:

\begin{equation}
\label{eq:bernoulii}
y_{i}(\theta) \sim \operatorname{Ber}\left(1 - e^{-\lambda_{i}(\theta)}\right).
\end{equation}

\subsection{ADCG-inspired reconstruction}
\label{Sec:ADCG_recon}
\begin{figure}
    \centering
    \includegraphics[width=1\linewidth]{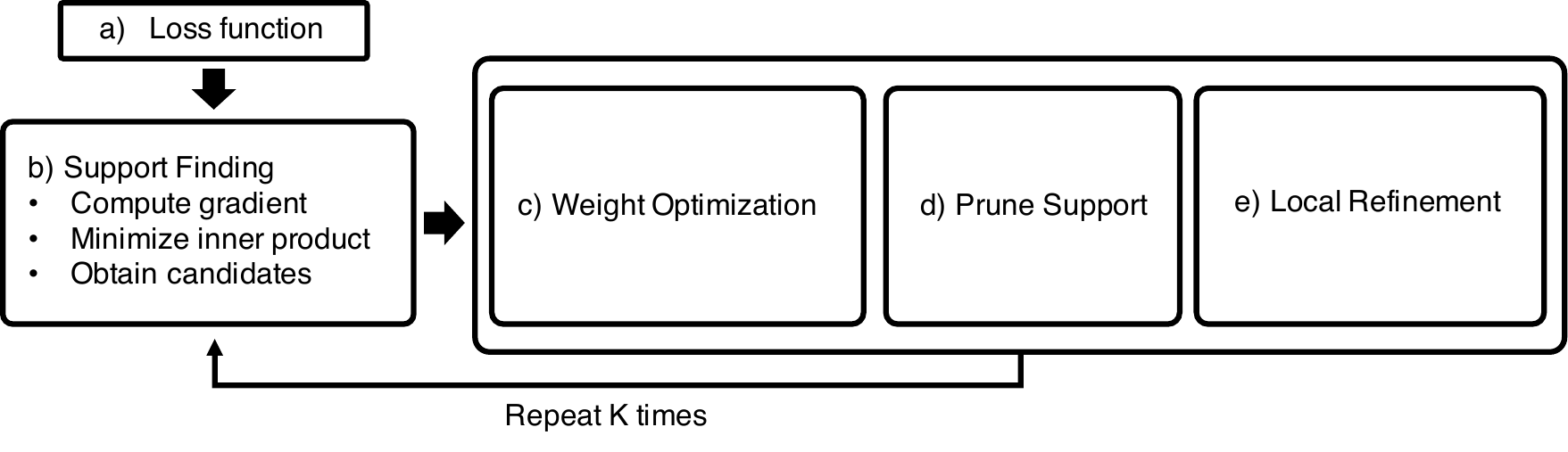}
    \caption{A flow diagram illustrating the main localization algorithm used in Section~\ref{Sec:ADCG_recon}.}
    \label{fig:ADCG_flow_diagram}
\end{figure}

To estimate the locations of simulated events from binary single-photon-sensitive sensor measurements, we employ a reconstruction algorithm inspired by the Alternating Descent Conditional Gradient (ADCG) framework~\cite{boyd2017alternating}. Figure~\ref{fig:ADCG_flow_diagram} illustrates the algorithm using a flow diagram and the procedure is described in detail in this section. 

We model the scintillation events $\mu$ as a nonnegative discrete measure over a continuous space $\Theta$:
\begin{equation}
\mu = \sum_{k=1}^{K}\alpha_k\delta_{\theta_k},
\label{eq:mu}
\end{equation}
where $\theta_k \in \Theta$ denotes the coordinates of the $k$-th event, $\alpha_k \ge 0$ is its expected detected photon count, and $\delta_{\theta_k}$ is the Dirac measure centered at $\theta_k$.
In particular, we primarily consider single-vertex ($K=1$) and intensity-limited double-vertex ($K=2$) cases in this paper, while retaining the measure-based formulation for potential extensions to multiple-vertex ($K>2$) events.

Given some scintillation events $\mu$, the expected photon rate at the $i$-th sensor pixel is modeled as
\begin{equation}
\lambda_i(\mu) = \int_{\Theta} P_i(\theta)d\mu(\theta),
\label{eq:forward_measure}
\end{equation}
where $P_i(\theta)$ is the normalized PSF defined in Eq.~\ref{eq:norm_psf}. 
By noting that $\mu$ consists of discrete events, as explained in Eq.~\ref{eq:mu}, the integral is reduced to
\begin{equation}
\lambda_i(\mu)=\sum_{k=1}^{K}\alpha_k P_i(\theta_k).\end{equation}

Consequently, by assuming a sparse and discrete source distribution, the continuous-domain integral formulation simplifies into a discrete and linear combination.

\paragraph{Loss function.}

Given the detection probability modeled in Eq.~\ref{eq:bernoulii} and binary single-photon-sensitive sensor measurement $\bm{y} \in \{0,1\}^m$, where $m$ denote the total number of activated pixels. Our loss function deploys negative log-likelihood (NLL) to evaluate how the probability function and measurement agree with each other and drive the optimization:

\begin{equation}
\mathcal{L}(\bm{\lambda)}
=
\sum_{i=1}^{m}
\Big[
(1-y_i)\lambda_i(\mu)
-
y_i\log\!\big(1-e^{-\lambda_i(\mu
)}\big)
\Big].
\label{eq:nll}
\end{equation}

\paragraph{Support finding (conditional gradient step).}

To identify the support, we minimize the directional derivative of the NLL defined in Eq.~\ref{eq:nll}. First, we compute the gradient $\bm{g}$ of the loss function with respect to the expected photon rate $\bm{\lambda}$, where each element $g_i$ is:

\begin{equation}
g_i = \frac{\partial \mathcal{L}}{\partial \lambda_i} = 1 - \frac{y_i}{1 - e^{-\lambda_i}}.
\end{equation}

The optimal support is determined by finding the continuous coordinate $\theta$ that minimizes the inner product with this gradient:

\begin{equation}
\theta^\star = \arg\min_{\theta \in \Theta} \langle \bm{g}, \bm{P}(\theta) \rangle.
\label{eq:cg_sparse}
\end{equation}

However, because the optical forward model is characterized by a finite set of experimentally calibrated measurements rather than an analytical, differentiable function, optimizing Eq.~\ref{eq:cg_sparse} directly over the continuous space $\Theta$ is intractable. In practice, we construct the forward model from a discrete grid of collected PSFs, denoted as $\Theta_{M} = \{\theta_1, \dots, \theta_M\}$. To approximate the continuous optimal support from these discrete measurements, we leverage the local shift-invariance of the optical system~\cite{ho2025differentiable}, which allows us to interpolate the forward model between calibration points. Specifically, for each reference PSF $\theta_m \in \Theta_M$, we define a continuous local neighborhood $\mathcal{N}_m$ within a bounding interval $\tau$:

\begin{equation}
\mathcal{N}_m = \{ \theta \in \Theta \mid |\theta - \theta_m| \le \tau \}
\end{equation}
and determine the optimal local translation within this region that minimizes the objective:
\begin{equation}
\theta_m^\star = \arg\min_{\theta \in \mathcal{N}_m} \langle \bm{g}, \bm{P}(\theta) \rangle, \quad \forall m \in {1, \dots, M}.
\label{eq:cg_step}
\end{equation}

The resulting set $\Theta_C = \{\theta^\star_1, \dots, \theta^\star_R\}$ forms the candidate support for the subsequent reconstruction steps.

\paragraph{Weight optimization.}

For a given candidate support set $\Theta_C$, we fix the localized support positions and optimize the corresponding emission weights. Specifically, for each $\theta_m^\star \in \Theta_C$, we obtain its weight $\alpha_m$ by solving

\begin{equation}
\min_{\alpha_m \ge 0} \mathcal{L}\left(\bm{\lambda}(\alpha_m; \theta_m^\star)\right).
\label{eq:weight_update}
\end{equation}

This optimization is implemented using Newton’s method~\cite{nocedal2006numerical}. By utilizing the second-order curvature of the NLL, Newton's method provides rapid, quadratic convergence toward the optimal photon counts. To ensure the physical validity of the weights, the updates are projected onto the non-negative constraint space $\alpha_m \ge 0$.

\paragraph{Support pruning.}

To enforce sparsity, we only keep the most likely scattering event in the candidate set $\Theta_C$. This is evaluated by computing the NLL for every combination of $(\alpha_C, \theta_C)$ that forms a candidate measure $\mu_C = \alpha_C \delta_{\theta_C}$:

\begin{equation}
\theta^\star_C = \arg\min_{\theta_C \in \Theta_C} \mathcal{L}\big(\bm{\lambda}(\mu_C = \alpha_C \delta_{\theta_C})\big).
\label{eq:support}
\end{equation}

\paragraph{Local refinement.}
The resulting optimal support $\theta^\star_C$ yields a coarse, on-grid estimate. To achieve off-grid precision, we perform a local refinement across both lateral and depth dimensions. Notably, while shift-invariance holds laterally, it does not extend to depth variations; these exhibit complex deformations that cannot be captured by simple spatial shifts or global scaling.

To address this, we employ the NLL of the PSFs as a proxy for depth refinement. Within a small neighborhood of the calibrated samples, the NLL varies smoothly along the $z$-axis; this allows us to model the loss trend using a parabolic function and determine the minimum in continuous space. We also account for the interdependency between depth variation and the effective magnification of the PSF. Once the depth estimate $\hat{z}$ is obtained, we apply compensatory scaling to the lateral positions, ensuring the final 3D localization aligns with the perspective geometry of the imaging system.

The process from $a$ to $e$ selects the individual event $\alpha_C \delta_{\theta_C}$ that minimizes the loss. This algorithm runs iteratively $K$ times to find the scattering events that best characterize the sparse photon measurements.

\subsection{Software Implementation}
All reconstructions are performed on a workstation equipped with 128~GB of RAM and an NVIDIA RTX~4060 GPU.
Due to memory limitations, the initial support search is performed using a downsampled version of the HDR PSF dictionary. 
Specifically, the calibrated PSF set containing 1210 PSFs is downsampled by a factor of two, resulting in a dataset of approximately 12.2~GB. 
The coarse support estimate obtained from the downsampled data is then used to select the corresponding full-resolution PSFs for final parameter estimation.
The full reconstruction pipeline is implemented in MATLAB with GPU acceleration.

\section{Results} \label{sec:results}

In this section, we test the reconstruction method on low photon count single-vertex and double-vertex images, which are produced by measured PSFs, to evaluate the system's performance for localizing single-scatter and double-scatter particle interactions in a scintillator volume. 

\subsection{Single scatter reconstruction} \label{Sec:single_scatter}

\begin{figure}
    \centering
    \includegraphics[width=1\linewidth]{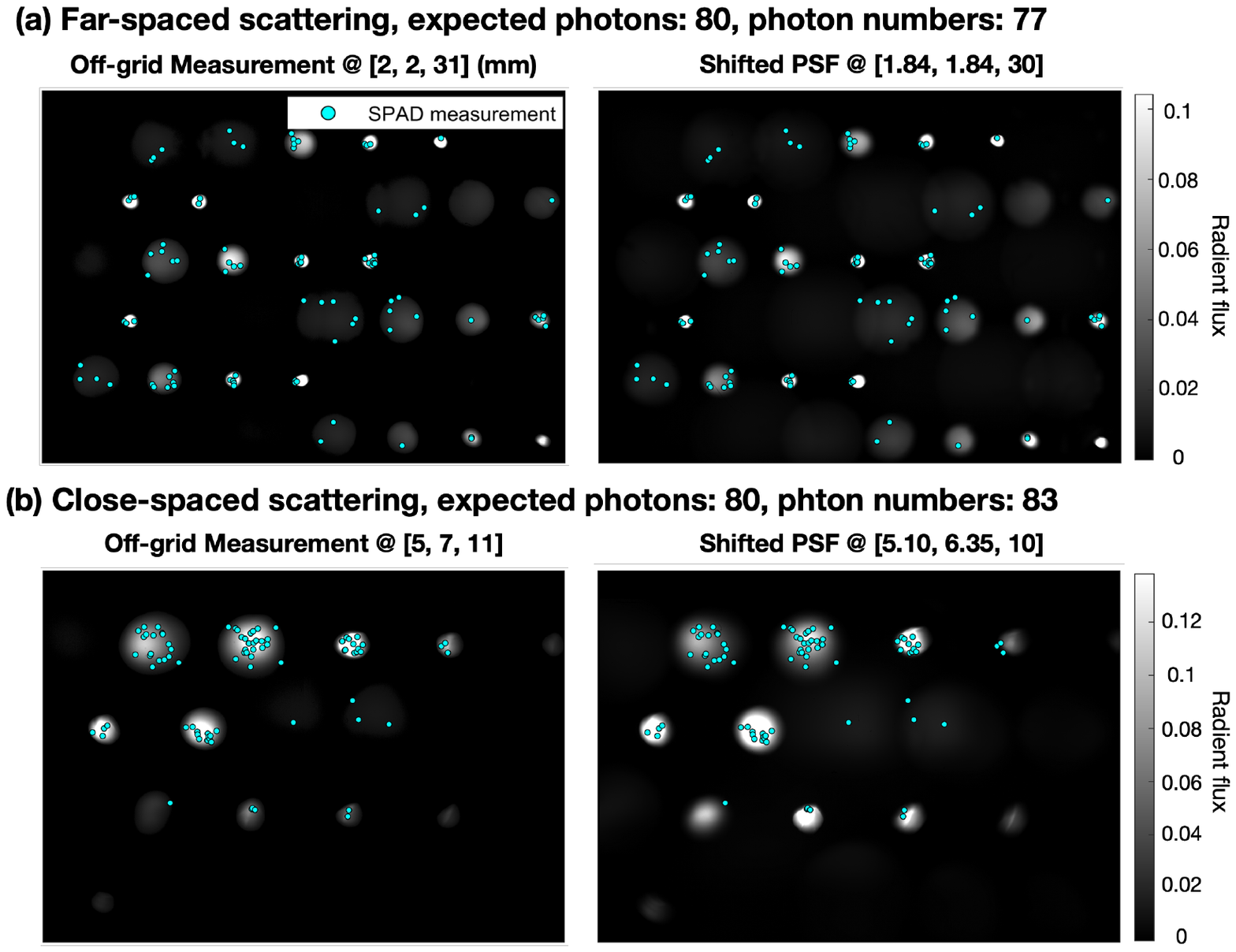}
    \caption{
        Illustration of support selection with local shift refinement. 
        Cyan markers denote binary single-photon-sensitive sensor detections, and grayscale intensity represents radiant flux (HDR measurement). Spatial coordinates are reported in millimeters (mm).
        \textbf{(a)} Far-spaced scattering case. An off-grid source at $[2,\,2,\,31]$ with 77 detected photons is shown (left). 
        The selected PSF is initialized from the nearest dictionary entry at $[1,\,1,\,30]$ and refined by applying an in-plane shift of $[\Delta u,\,\Delta v] = [1.84,\,1.84]$ (right).
        \textbf{(b)} Close-spaced scattering case. An off-grid source at $[5,\,7,\,11]$ with 83 detected photons is shown (left). 
        The selected PSF at $[5,\,5,\,10]$ is refined using a shift of $[\Delta u,\,\Delta v] = [0.10,\,1.35]$ (right).
    }
    \label{fig:recon_visual}
\end{figure}

Figure~\ref{fig:recon_visual} visualizes the support selection and refinement process using the sensor image for the ADCG-based reconstruction of single-vertex events. Here we show representative results for two off-grid single-scatter events, the true positions of which do not coincide with any calibration grid points used in the reconstruction, with 80 expected photons. 

In the left column, cyan markers denote sparse binary photon detections generated by measured PSFs using the single-photon-sensitive sensor model, while the grayscale image shows the corresponding on-grid HDR PSF that best matches the test data. 
The right column shows the refined PSF obtained by locally shifting the selected calibration PSF around the best-match grid point to further improve data-model agreement, as defined in Eq.~\ref{eq:cg_step}. 
In addition, the top row example illustrates a test source position far (X~mm) from the optics, and the bottom one shows that for a nearby source (Y~mm)
In both examples, the shifted PSFs closely align with the off-grid binary measurements, demonstrating the effectiveness of the ADCG-based support selection and local refinement procedure.

\begin{figure*}
    \centering
    \includegraphics[width=\linewidth]{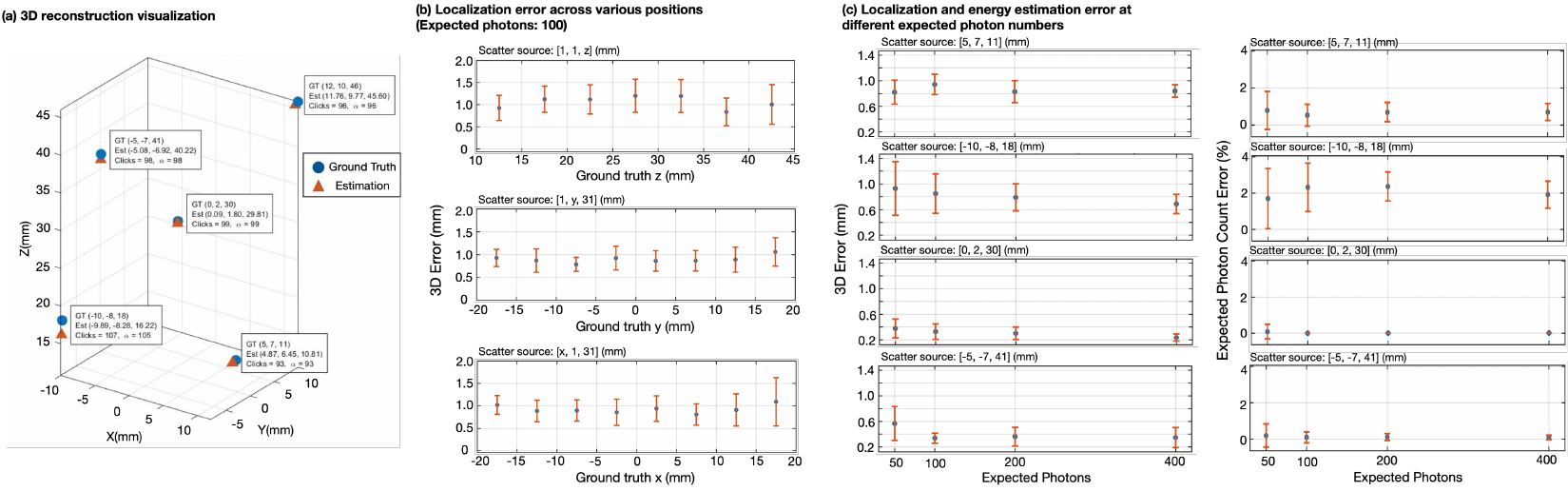}
    \caption{Reconstruction performance under varying photon conditions.
    \textbf{(a)} 3D visualization of ground truth (blue circles) and reconstructed positions (triangles) at 100 expected photons. The total number of detected clicks is derived from binary single-photon-sensitive sensor measurements, and $\alpha$ denotes the estimated expected photon count.\textbf{(b)} 3D localization error at 100 expected photons over 25 trials per test point. Rows correspond to varying one spatial dimension while fixing the other two: top ($z$), middle ($y$), and bottom ($x$).\textbf{(c)} Reconstruction performance at different expected photon levels (50, 100, 200, and 400). Each condition includes 25 trials per position. The single best and single worst estimates are excluded.}
    \label{fig:result}
\end{figure*}

We next present qualitative and quantitative reconstruction results in Fig.~\ref{fig:result}. 
Fig.~\ref{fig:result}(a) shows the reconstruction results for five representative single-vertex source locations with 100 expected photons. 
The realized binary photon detection counts for each test are labeled as \textit{Clicks}, and the estimated expected photon counts are denoted as $\alpha$. 
Blue markers indicate the ground-truth locations, while orange triangles represent the reconstructed positions. 
The estimated positions closely match the ground truth across the tested locations, and the estimated expected photon counts are also consistent with the detected photon counts in all examples.

Unless otherwise noted, each data point in the quantitative result is computed from 25 trials, with the single best and single worst estimates excluded before calculating the reported mean and standard deviation. 
This trimming reduces the influence of extreme photon realizations while preserving the overall reconstruction trend.
Fig.~\ref{fig:result}(b) summarizes the 3D localization error as a function of test source positions with 100 expected photons. 
In each row, we vary one coordinate of the test source position (from top to bottom, $z$, $y$, and $x$, respectively), while holding the other two coordinates fixed. 
The mean 3D localization error remains near 1~mm across most of the tested field of view, with a relatively small standard deviation. 
Notably, the error remains stable over the tested axial range, as CRLB results showed in Fig.~\ref{fig:crlb}, confirming that the multifocal plenoptic design provides relatively homogeneous depth encoding for 3D localization. 
Larger errors are observed at off-axis lateral positions near the edge of the field of view. 
This increase is partly due to the finite aperture of the lens array: the active lens-array aperture is approximately $3~\mathrm{cm} \times 4~\mathrm{cm}$, which is smaller than the $5~\mathrm{cm} \times 5~\mathrm{cm}$ scintillator cross section.
As a result, photons emitted from off-axis positions are collected over a more limited angular range, reducing localization accuracy near the edge of the volume.
A detailed per-axis error analysis is provided in Fig.~S1 of the Supplemental Material~\cite{supplemental}.

Fig.~\ref{fig:result}(c) shows the 3D localization error and the expected photon-count estimation error as a function of expected photon counts.
As expected, both errors increase as the expected photon count decreases, reflecting the challenges in light-starved event localization. 
The reconstruction is limited by the system when more than 100 expected photons are detected. 
At 50 expected photons, occasional large-error trials occur because the realized binary detections may become too sparse to adequately sample the underlying photon distribution.

\subsection{Double scatter reconstruction}
\label{sec:double scatter}
\begin{figure}
    \centering
    \includegraphics[width=1\linewidth]{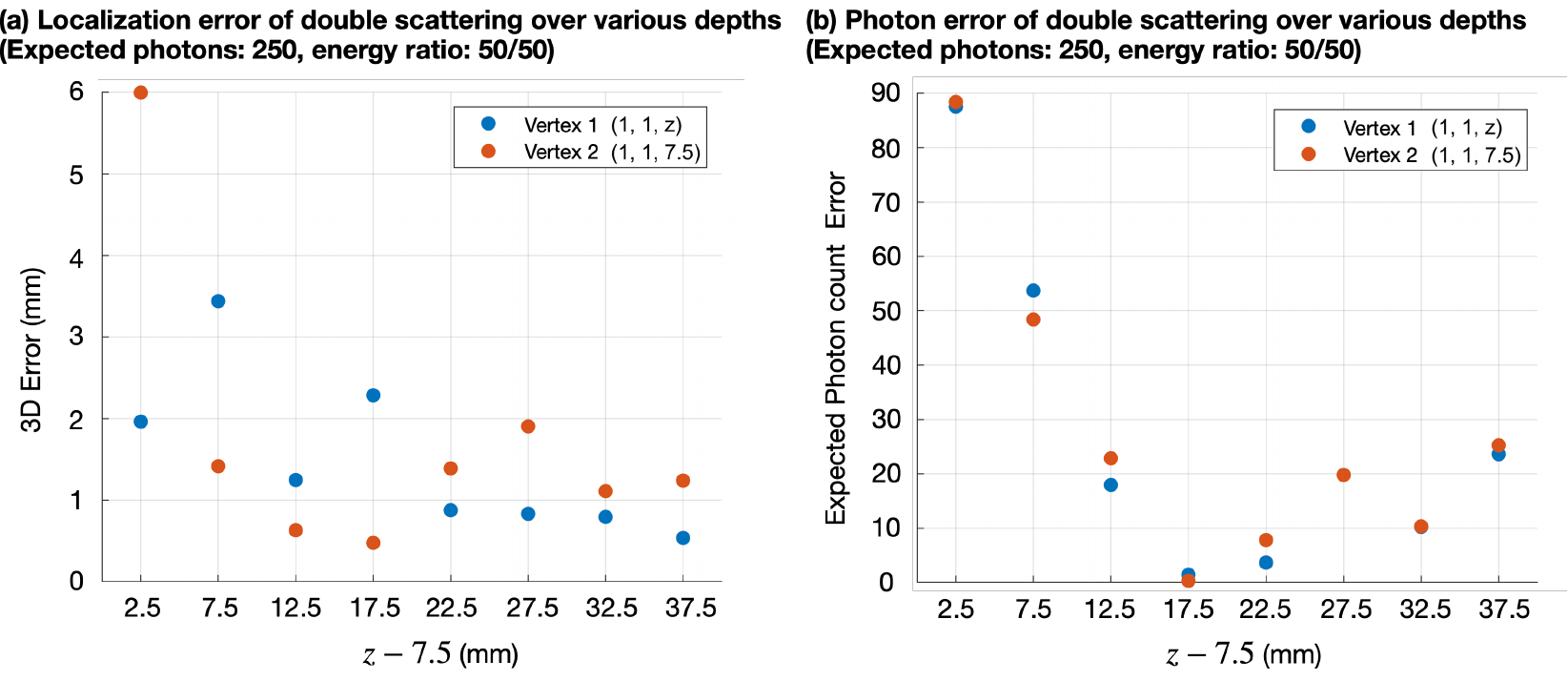}
    \caption{
    \textbf{(a)} 3D localization error and \textbf{(b)} absolute expected photon count error as a function of axial separation between two vertices. 
    The horizontal axis denotes the difference $z - 7.5$ (mm), while both vertices share the same lateral position $(x,y) = (1,1)$ mm. 
    Blue markers correspond to vertex~1 at $(1,1,z)$ mm, and orange markers correspond to vertex~2 fixed at $(1,1,7.5)$ mm. 
    All measurements are generated with 250 expected photons and a 50/50 energy ratio between the first and second vertices.
    The single best and single worst estimates are excluded
    }
    \label{fig:doublescattering}
\end{figure}

The reconstruction framework is further tested with double-scatter events, where the sparsity level in the ADCG-based algorithm is set to $K=2$, allowing the locations and photon weights of two sources to be estimated. However, the addition of a second vertex significantly increases the complexity of the reconstruction and its interpretation, both of which depend strongly on the relative separation and scintillation intensity of the two vertices. Therefore, in this work, we only study the simple case when the two light sources share equal expected photon weights. A more comprehensive evaluation of the system's performance for general multi-scatter event reconstruction will be carried out in a future dedicated study. 

For a double-scatter event, the sensor measurement would consist of superimposed optical responses from two sources. In this test, we generate double-scatter images by linearly adding two downsampled single-vertex images corresponding to two selected source locations. 
The overlap of their optical responses render the inverse problem difficult, especially when the two sources share similar locations and thus produce similar PSFs. 
To evaluate a particularly challenging case, we place both vertices at the same lateral position, $(x,y)=(1,1)$~mm, so that their responses primarily differ along the axial direction. 
Specifically, vertex~1 is placed at $(1,1,z)$~mm with varying depth, while vertex~2 is fixed at $(1,1,7.5)$~mm, with each vertex producing an expected photon count of 125. 
The labels vertex~1 and vertex~2 are used only to distinguish the two sources and do not indicate temporal ordering.

The reconstruction results for this case study are summarized in Fig.~\ref{fig:doublescattering}. 
As expected, reconstruction is most difficult when the two vertices are close in depth. 
In this regime, their PSFs overlap strongly and become nearly degenerate, leading to larger errors in both localization and photon-weight estimation for individual sources. 
As the axial separation increases, the two depth-dependent responses become more distinguishable, generally reducing both localization and photon-weight errors.
This observation confirms that the multifocal plenoptic design produces useful information for separating multiple sources, especially when their optical responses are sufficiently distinct.

These results provide an preliminary demonstration of double-scatter reconstruction capabilities using the proposed optical design and reconstruction framework. 
Although this study is limited to equal photon weights and a fixed lateral configuration, it provides a feasible path toward multi-source localization in light-starved scintillation detectors. 
Future work will investigate more general multiscatter event reconstruction with varying spatial and photon weight configurations at high statistics, and evaluate the sensitivity of such a system for various applications.

\section{Discussion} \label{sec:discussion}
This work suggests that plenoptic imaging systems can be optimized for 3D event localization in light-starved environments with a compact form factor. 
A central design challenge is that the lens array and photosensor need to be placed as close as possible to the scintillator to maximize light collection efficiency, which is critical for scintillation imaging where the total number of available photons is small. 
However, a small object-lens separation can introduce excessive defocus and degrade image sharpness across much of the target volume. 
The plenoptic prototype system design presented here addresses this challenge by adopting an unconventional design without a main lens and by carefully selecting a variety of focal lengths for the MLA, as guided by a CRLB analysis.
The system enables a short standoff distance while maintaining sufficient optical encoding for 3D reconstruction within the target volume.

Alternative approaches for 3D spatial reconstruction in single-volume scintillator detectors, such as centroiding~\cite{braverman2018single,harvey2021development} and coded apertures~\cite{folsom2020compact,ziock2019real},  have also been explored in light-starved environments. Here we provide a brief analysis of the differences between our plenoptic techniques and these established methods.

The centroiding technique couples a single-volume scintillator to pixelated photosensors on multiple faces of the scintillator~\cite{braverman2018single,harvey2021development}. Because scintillation light is emitted isotropically, depth information is encoded in the recorded sensor hit pattern. However, the depth encoding is fundamentally weak, making axial inference challenging. As a result, detectors with centimeter-scale dimensions naturally approach centimeter-scale depth resolution limits, and the axial resolution also degrades rapidly as detector depth increases and photon statistics decrease.

Coded aperture techniques introduce a thin mask between the scintillator and the pixelated photosensors. This mask encodes interaction positions into structured intensity patterns at the sensor plane~\cite{folsom2020compact,ziock2019real}, enabling improved depth sensitivity relative to centroiding while largely preserving lateral resolution. However, amplitude-based encoding requires a fraction of scintillation photons to be blocked by the mask and attenuates the detected signal strength. In addition, the observed photon signal is distributed across a large number of sensor pixels, increasing sensitivity to readout noise and potentially limiting performance in photon-starved regimes.

Plenoptic imaging obtains lateral and depth sensitivity through phase-based optical encoding that redistributes rather than absorbs incident photons. The MLA, as a compact implementation of plenoptic imaging, preserves photon throughput in contrast to amplitude-mask approaches such as coded apertures. In low-photon regimes, this photon-efficient property can be particularly advantageous. Moreover, a fraction of the light is concentrated onto localized regions of the sensor array, reducing the noise-compounding effect encountered in coded-aperture and centroiding techniques.

For many radiation imaging applications, such as neutron double-scatter cameras~\cite{braverman2018single}, Compton imagers~\cite{h3d_2019}, and topology-based particle identification techniques~\cite{liquido_2025}, mm-scale 3D spatial resolution is desirable, as interaction separations are typically on the cm scale. Our results indicate that plenoptic systems can achieve mm-scale 3D localization for both single-vertex and double-vertex events with a compact form factor. We also observe that axial (z) resolution tends to dominate the overall 3D localization performance. As a result, a strategy to improve spatial resolution is to employ two orthogonally oriented plenoptic systems for the same scintillator volume, such that the lateral sensitivity of one system complements the axial sensitivity of the other. 
We comment that our work represents a preliminary study of plenoptic design optimization for light-starved imaging of sparse sources, while a vast design parameter space remains unexplored, so future work can potentially achieve much improved 3D resolution using a single plenoptic system.

\section{Conclusion} \label{sec:conclusion}
\nocite{*}
The PRISM architecture demonstrates that a multi-focal plenoptic design with a single-photon-sensitive sensor array can enable accurate 3D localization in the light-starved regime while maintaining a wide field of view. 
The system is designed to preserve photon efficiency and demonstrates stable reconstruction performance under low photon counts using data-driven PSF calibration. 
In addition, the use of CRLB analysis provides theoretical insight and serves as a guideline for selecting the multi-focal lens configuration. 
Together, these results establish PRISM as a promising approach for photon-limited volumetric imaging.

Importantly, we foresee an improved system performance with further optimized implementations of this imaging concept. 
First, the optical design has not been systematically optimized, with a large design space remaining to be explored. Potential improvements include the lens arrangement, the selection of focal lengths for each lens, and the adoption of a multi-focal, multi-size MLA design that should improve energy uniformity by constraining the $f$-number across all lenses.
Second, incorporating additional priors or learning-based components into the reconstruction pipeline may further improve reconstruction accuracy and robustness.
Future research directions also include the dedicated investigation of multi-scatter event reconstruction and experimental tests with a physical scintillator to study the complexity of scintillation processes, including spectral variation, temporal dynamics, and volumetric emission. 

Overall, PRISM provides a promising foundation for photon-efficient 3D imaging in challenging conditions, with substantial opportunities for further improvement through joint optical and computational design.

\section{Acknowledgement}
The authors wish to acknowledge Timothy Classen and Nathaniel Bowden for assistance in securing experimental space and providing equipment used for prototype testing. We thank Jeremy Lusk for useful discussions on the friction-based lens holder design. The work of Andreas Velten, Forrest Peterson, Alex Bocchieri, and David Parra was partially supported by the Consortium for Enabling Technologies and Innovation-2.0. This work was supported by the U.S. Department of Energy National Nuclear Security Administration and Lawrence Livermore National Laboratory [Contract No. DE-AC52-07NA27344, release number LLNL-JRNL-2020946]. This work was supported by the U.S. Department of Energy Office of Defense Nuclear Nonproliferation Research and Development.

\bibliographystyle{apsrev4-2}
\bibliography{bibfile}

\end{document}


\title{Supplemental Material for Plenoptic imaging of particle interactions in scintillation detectors}
\maketitle

\section{Additional Reconstruction Results}
Fig.~S1 provides a detailed decomposition of the 3D localization error presented in the main text. Specifically, we report the reconstruction error along each spatial coordinate (x, y, z), as well as the combined 3D error, as individual dimensions are varied while holding the others fixed.

While the main paper reports only the overall 3D localization error, this figure reveals the underlying contributions from each axis in the main manuscript Fig.~7. The lateral errors (x and y) remain relatively small and stable across the field of view, whereas the axial (z) error exhibits higher variance. This behavior is consistent with the reduced depth sensitivity of the optical system under photon-limited conditions.

Additionally, increased variability is observed near the edges of the field of view, where the system response becomes less informative. Overall, this breakdown confirms that the millimeter-scale 3D accuracy reported in the main paper is achieved through strong lateral localization, with axial estimation representing the dominant source of error.
\begin{figure}[!htbp]
    \centering
    \includegraphics[width=\linewidth]{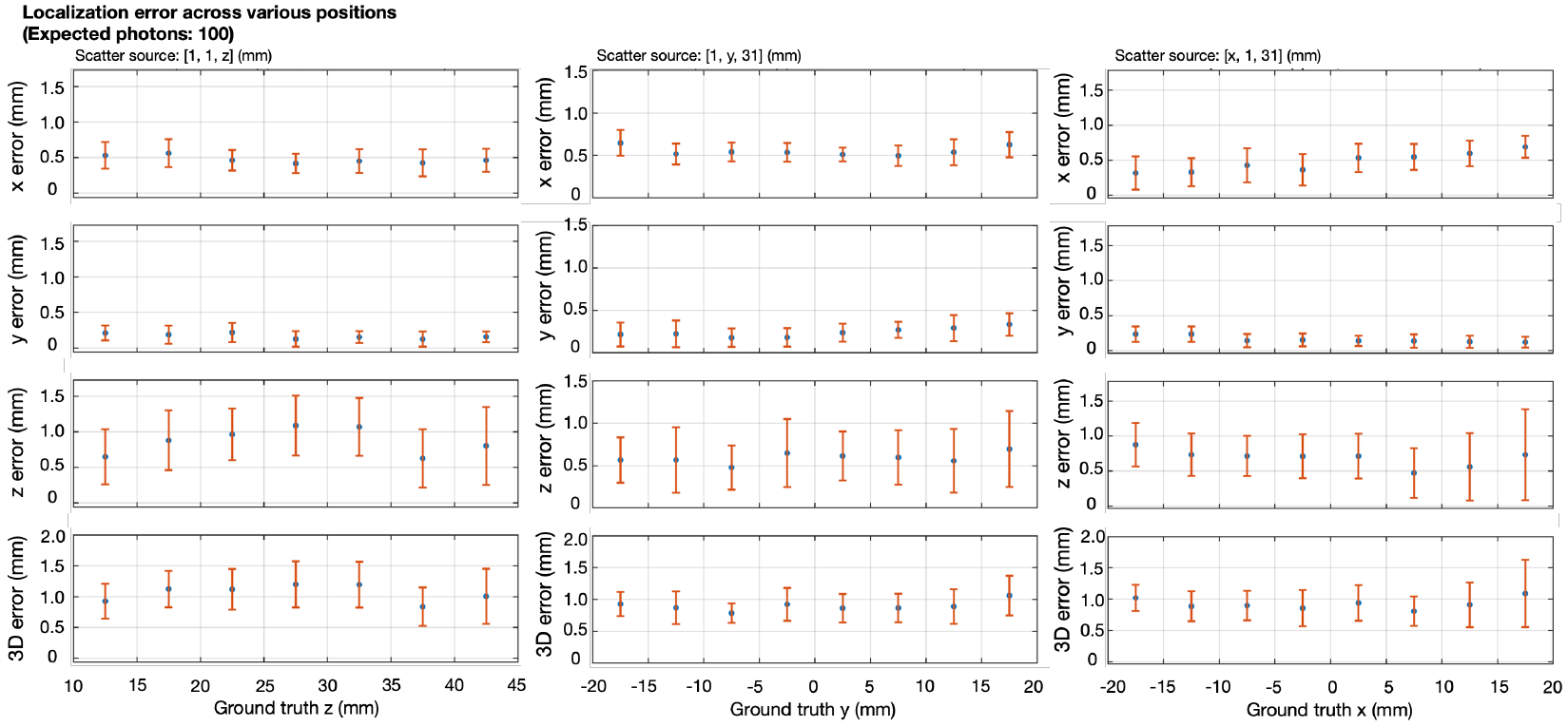}
    \caption{
    Breakdown of localization error across spatial positions for a representative scatterer at $(1,1,31)$ mm. The figure decomposes the total 3D error into per-axis components (x, y, z) as individual spatial coordinates are varied while the remaining coordinates are held fixed. Error bars denote standard deviation over multiple trials. The results show that the overall 3D error reported in the main text is primarily driven by larger variance in the axial (z) direction, while lateral errors remain relatively small and stable across the field of view.
    }
    \label{fig:extra recon result}
\end{figure}